\documentclass[letterpaper,twocolumn,10pt]{article}
\usepackage{usenix-2020-09}

\usepackage{tikz}
\usepackage{amsmath}
\usepackage{setspace}

\usepackage[linesnumbered,ruled,vlined]{algorithm2e}

\SetCommentSty{mycommfont}
\SetKwInput{KwInput}{Input}                
\SetKwInput{KwOutput}{Output}              
\SetKwFor{Block}{\(\blacktriangledown\)}{}{}
\SetKwFunction{hide}{CAMOUFLAGE}
\SetKwFunction{blast}{BLAST}
\SetKwFunction{screen}{SCREENER}
\SetKwFunction{score}{SCORE}
\SetKwFunction{extend}{EXTEND}
\SetKwFunction{merge}{MERGE}
\SetKwProg{Fn}{Function}{:}{}

\graphicspath{ {./images/} }

\newcommand{\bestmatch}{\emph{Best Match}}
\newcommand{\sat}{{SoC}}
\newcommand{\nsa}{{NSoC}}
\newcommand{\rsap}{{RDI}}
\newcommand{\osaone}{{SoCO1}}
\newcommand{\osatwo}{{SoCO2}}
\newcommand{\genothreat}{{GenoTHREAT}}
\newcommand{\gtp}{{GTP}}
\newcommand{\ged}{{Gene Edit Distance}}

\newcommand{\cas}{{Cas9}}

\usepackage{times}
\usepackage{multirow}
\usepackage[utf8]{inputenc}
\usepackage{graphicx}
\usepackage{authblk}
\usepackage{url}
\usepackage{color}
\usepackage{xcolor}
\usepackage{amssymb}
\usepackage{flushend}

\usepackage[T1]{fontenc}
\usepackage{pifont}

\newtheorem{definition}{Definition}
\newtheorem{example}{Example}

\renewcommand{\paragraph}[1]{\noindent\textit{\textbf{#1.~}}}

\usepackage{enumitem}
\setlist{nolistsep,leftmargin=*}

\begin{document}

\date{}

\title{\Large \bf Cyberbiosecurity: DNA Injection Attack in Synthetic Biology
}

 \author{
{\rm Dor Farbiash}\\
Jusidman Science Center for Youth at Ben-Gurion University of the Negev
\and
{\rm Rami Puzis}\\
Cyber Security Research Center and 
Software and Information Systems Engineering Department at
Ben-Gurion University of the Negev
} 

\maketitle
\begin{abstract}

Today arbitrary synthetic DNA can be ordered online and delivered within several days.  
In order to regulate both intentional and unintentional generation of dangerous substances, most synthetic gene providers screen DNA orders.  
A weakness in the Screening Framework Guidance for Providers of Synthetic Double-Stranded DNA allows screening protocols based on this guidance to be circumvented using a generic obfuscation procedure inspired by early malware obfuscation techniques.
Furthermore, accessibility and automation of the synthetic gene engineering workflow, combined with insufficient cybersecurity controls, allow malware to interfere with biological processes within the victim's lab, closing the loop with the possibility of an exploit written into a DNA molecule presented by Ney et al. in USENIX Security'17.   
Here we present an end-to-end cyberbiological attack, in which unwitting biologists may be tricked into generating dangerous substances within their labs.
Consequently, despite common biosecurity assumptions, the attacker does not need to have physical contact with the generated substance.  
The most challenging part of the attack, decoding of the obfuscated DNA, is executed within living cells while using primitive biological operations commonly employed by biologists during in-vivo gene editing.   
This attack scenario underlines the need to harden the synthetic DNA supply chain with protections against cyberbiological threats. 
To address these threats we propose an improved screening protocol that takes into account in-vivo gene editing.

\end{abstract}

\vspace{-0.4cm}
\section{Introduction}
\vspace{-0.2cm}
Synthetic biology is an emerging bioengineering technology that plays a significant role in personalized medicine, pharmaceutical manufacturing, etc.\cite{heinemann2006dnamedications,lienert2014synthetic}.
Rapid development of biological systems is supported by large online libraries of genes~\cite{kamens2015addgene,yamazaki2017biobrick,benson2012genbank}, as well as integrated development environments (IDEs) and gene compilers for efficient gene coding~\cite{czar2009writing}.  
Currently, the software stack used to develop synthetic genes is loosely secured (Section~\ref{sec:vuln}), potentially allowing the injection of rogue genetic information into biological systems by a malware. 

DNA synthesis companies, which produce and ship the DNA sequences, are an important element of the growing synthetic biology market.
Synthetic DNA is available in multiple ready-to-use forms, such as a circular DNA molecule called a plasmid (Section~\ref{sec:dna-background}). 
A synthesized plasmid can be inserted into an organism by following a simple biological protocol, after which it can start producing proteins~\cite{mandecki1990totally}.  
Many bioengineering tools are now easily accessible by 
do-it-yourself (DIY) biology enthusiasts. 

The products of biological systems can be extremely dangerous substances, such as toxins or synthetic viruses~\cite{cello2002chemical,koblentz2017novo}.
DNA sequences producing dangerous viruses may be as small as 1.7kbp~\cite{huang2010evolution}; natural and synthetic toxins may be constructed from less than 100bp~\cite{lipps2000small}.     
The dual use of synthetic biology as a powerful technology for the benefit of mankind, and as a potential weapon, is a long-standing issue~\cite{national2018biodefense}. 
The dangers of synthetic biology are manifold, and they require rigorous security controls~\cite{gronvall2019synthetic,west2020crispr}.
One such control is the U.S. Department of Health and Human Services' (HSS) Screening Framework Guidance for Providers of Synthetic Double-Stranded DNA (HHS guidelines)~\cite{hhsscreening}. 
We discuss state-of-the-art DNA screening in Section~\ref{sec:screening}. 
Biosecurity researchers agree that constant improvement of DNA screening methodologies is required to prevent both bioterrorists, and careless enthusiasts, from generating dangerous substances in their labs~\cite{national2018biodefense,kobokovich2019genesynthesissecurity}.  
Legislation is keeping pace. Recently California obliged all customers to order synthetic genes from companies that perform gene screening~\cite{west2020california}.

Understanding the impact of cyber threats on biosecurity is extremely important. 
Cyber attacks are considered a potential threat to the security and privacy of genomic data~\cite{huang2015genoguard,caswell2019defending} and the analysis of genetic material~\cite{fayans2020cyber}. 
A recent biodefense report~\cite{national2018biodefense} mentions cyber threats with respect to identifying potential targets and the development of bioweapons.   
Diggans and Leproust~\cite{diggans2019next} highlight the value of cybersecurity methods applied in the domain of biosecurity.    
Ney at el.~\cite{ney2017computer} have even suggested an attack where an exploit written in a DNA molecule may subvert a vulnerable sequencing machine.

However, the potential of delivering a harmful biological agent through cyberspace has not yet been considered.  
It is currently believed that a criminal needs to have physical contact with the substance to produce and deliver it. 
However, malware could easily replace a short sub-string of the DNA on a bioengineer's computer with a toxin producing sequence.   
Order screening by synthetic DNA providers is the most effective line of defense against such attacks.  
Unfortunately, the screening guidelines have not been adapted to reflect recent developments in synthetic biology and cyberwarfare.

In this paper we discuss two main attack vectors:  

\noindent(1) A traditional biosecurity vector where an adversary orders a dangerous synthetic DNA: 
\begin{itemize}
    \item We show that a weakness in the HHS guidelines can be exploited by the attacker to avoid detection by obfuscating the malicious DNA (Section~\ref{sec:seqobfuscation}). 
    \item The proposed Gene Edit Distance (Section~\ref{sec:ged}) can efficiently detect sequences that can be decoded into malicious DNA within living cells.
    \item A benchmark dataset of obfuscated DNA sequences demonstrates the superiority of Gene Edit Distance over a rigorous implementation of the HHS guidelines (Section~\ref{sec:eval}). 
\end{itemize}

\noindent(2) A cyberbiological attack vector where malware on the biologist's computer interferes with biological processes: 
\begin{itemize}
    \item We discuss the attack surface of a synthetic biology pipeline (Section~\ref{sec:vuln}) and highlight the overlooked impacts malware may have on biological processes (Section~\ref{sec:impacts}).  
    \item We demonstrate an end-to-end scenario where a malware within a biological lab tricks the biologist victim into producing a substance of the attacker's choice.  
    This scenario closes the loop with a potential attack propagating from DNA to cyberspace~\cite{ney2017computer} and stresses the importance of designing adversary-resilient biological protocols. 
    \item Simple mitigation steps applied by DNA providers (Section~\ref{sec:discussion}) may significantly reduce the threat of a cyber attack propagating to the biological space.     
\end{itemize}

\vspace{-0.2cm}
\section{Background and related work}
\vspace{-0.2cm}
\label{sec:background}
This section provides the most important information required in subsequent sections, where we describe potential cyberbiological threats and their mitigation. 
Readers familiar with the basics of protein synthesis and the CRISPR-Cas system are encouraged to skip Section~\ref{sec:dna-background}.  
Important terms used later in the paper are highlighted in \textbf{bold} face. 

\vspace{-0.2cm}
\subsection{An introduction to DNA editing for cybersecurity experts}
\label{sec:dna-background}
Genetic information in living cells is encoded in sequences of \textbf{nucleotides} called DNA. 
Nucleotides are commonly denoted by four letters, C,G,A, and T, corresponding to four nucleobases; cytosine, guanine, adenine, and thymine, respectively. 
In a double-stranded DNA (\textbf{dsDNA}) molecule, nucleobases on the opposite strands are bound together, C with G and A with T. 
Thus, dsDNA is sometimes seen as a sequence of \textbf{base pairs (bp)}. 
The fees charged by gene synthesis companies for synthetic DNA orders are typically based on the number of base pairs. 
The orders are commonly delivered in the form of cyclic dsDNA molecules called \textbf{plasmids} which are very stable and can be replicated within a living cell. 
The price for a plasmid can be as low as five cents per base pair.  
Once provided with a plasmid, one can employ a \textbf{sequencing} procedure to get the string representation of the DNA molecule. 
Companies that produce synthetic DNA usually also provide sequencing services. 

A complement of a DNA sequence is formed by replacing every occurrence character with its pair ($C\leftrightarrow G, A\leftrightarrow T$). 
\textbf{A reverse complement} of a DNA sequence, an operation commonly used in gene design, is formed by reversing the complement of a DNA sequence. For example, the complement of GGCA is CCGT and its reverse complement is TGCC. 

Along with other cell functions, DNA encodes proteins. 
First, an RNA sequence is generated from a DNA region surrounded by special sequences called \textbf{promoters} and \textbf{terminators}. 
An RNA molecule contains the same information as the respective DNA, but it is short-lived. 
In computer terms it can be compared to volatile memory, whereas DNA would be considered persistent storage. 
RNA containing special sequences called \textbf{ribosome binding sites} can be transformed by a ribosome into a sequence of amino acids. 
Every three nucleotides form one \textbf{amino acid}, but different triplets (a \textbf{codon}) may form the same amino acid.
Translating amino acids back to triplets of nucleotides is called \textbf{reverse translation}. 
The choice of which triplets are optimal depends on the organism in which the DNA is expressed.    
There are 20 amino acids.  
Short sequences of amino acids are called \textbf{peptides}, while long sequences are called \textbf{proteins}. 
Both perform various functions within a living cell.

The clustered regularly interspaced short palindromic repeats (\textbf{CRISPR}) complex is a part of the bacterial immune system that was adapted by bioengineers to perform precise DNA editing in live biological systems. 
The most common DNA editing system consists of a \textbf{\cas} protein and a \textbf{guide RNA sequence (gRNA)}. 
In this article we will use the term CRISPR to refer to the DNA system consisting of \cas~and gRNA. 
The \cas~protein performs a cut in a dsDNA molecule at specific locations called protospacer adjacent motifs (\textbf{PAMs}). 
The gRNA contains a short replica of the region following the PAM that needs to be cut by \cas. 
In computer terms, gRNAs can be regarded as pointers in an associative memory.
For gRNA creation, the DNA should contain a promoter, a copy of the \textbf{gRNA target site}, and a terminator; collectively these are referred to as a \textbf{gRNA scaffold}.

A dsDNA that was cut by a CRISPR can repair itself. 
Such a repair process is error prone and can produce mutations at the cut point. 
If such a mutation results in production of a different amino acid during protein formation, the protein may become non-functional (a.k.a. a gene \textbf{knockout}). 
Precise repairs of the cut DNA can be performed using a process known as \textbf{homology directed repair (HDR)}. 
To activate HDR, the cell should contain a DNA sequence that repeats the sequence of nucleotides to the left and right of the cut point (left and right arms of the HDR template respectively) and a small number of nucleotides that can be inserted between them at the cut point. 
HDR can also correct a few small mutations close to the cut point. Using CRISPR and HDR it is possible to remove and replace long DNA fragments, a process known as \textbf{knock-in}~\cite{wang2015large,pohl2016crispr}. 
The success rate of knock-in may vary significantly, from 2-3\% up to almost 80\%~\cite{hui2018comparing,kumita2019efficient}.

\vspace{-0.2cm}
\subsection{Sequence alignment}
\label{sec:blast}
Next we discuss sequence alignment, which plays a significant role in bioengineering and in biosecurity.    
The \textbf{Basic Local Alignment Search Tool (\blast)}~\cite{altschul1990basic} is the first of a long series of algorithms developed for aligning sequences of nucleotides or amino acids.   
\blast algorithms are optimized both for speed and for searching large databases of sequences. 
Let $q$ denote a \textbf{query sequence} and $t$ denote a subject sequence (also called a \textbf{target sequence}). 
Let $q[i]$ denote the $i$'th character in $q$.   
\blast operates by matching n-grams---short sequences of letters (words)---and extending these matches to form local \textbf{alignments} between the sequences.

Provided a query sequence, \blast returns a set of target sequences similar to the query sequence and a set of alignments (also called ranges) for each target sequence.
Let $\mathcal{A}_{q,t}$ be a set of such alignments between $q$ and $t$ found by \blast. 
Every alignment in $\alpha\in \mathcal{A}_{q,t}$ maps a range of character positions (a substring) in $q$ to a range of character positions (a substring) in $t$, such that any two successfully aligned character positions $i>j$, $\alpha(i)>\alpha(j)$.
Although $\alpha$ is not a function, we use the terms domain ($dom(\alpha)$) and image ($img(\alpha)=a(dom(\alpha))$) to denote the respective substrings. 
Let $\alpha^{-1}$ denote the inverse alignment, such that $dom(\alpha^{-1})=img(\alpha)$ and $img(\alpha^{-1})=dom(\alpha)$.

The score of an alignment is computed based on the number of \textbf{matched} characters  ($M=|\{i:q[i]=t[\alpha(i)]\}|$); the number of \textbf{mismatched} characters ($MM=|\{i:q[i]\neq t[\alpha(i)]\}|$); the number of \textbf{gaps opened} in both the query and the target sequences $G=|\{i:\alpha(i-1)\neq\bot\wedge\alpha(i)=\bot\}|+|\{i:\alpha^{-1}(i-1)\neq\bot\wedge\alpha^{-1}(i)=\bot\}|$; and the total \textbf{extent} of the gaps ($GX=|\{i:\alpha(i)=\bot\}|+|\{i:\alpha^{-1}(i)=\bot\}|$), where $\bot$ means that the argument character is not aligned. 
There is a reward ($rm$) for every matching character and penalties for mismatching characters ($pmm$), gap opening ($pgo$), and gap extension ($pgx$). 
The reward and penalties are configurable. 
The score of an alignment is: 
$$
Score = rm\cdot M - pmm\cdot MM - pgo\cdot G - pgx\cdot GX
$$
The fraction of $q$ characters successfully aligned is called \textbf{query coverage}:
$    QC(\alpha) = \frac{MM}{\left|q\right|}$.
The \textbf{percent identity} is computed from the sizes of the domain, the image, and the number of successfully mapped characters:
$    PI(\alpha) = 2\cdot MM/(\left|dom(\alpha)\right|+\left|img(\alpha)\right|) $

\begin{example}
Consider the following alignment:
$\alpha=\{(1,1), (2,2), (3,\bot), (4,3), (\bot,4), (\bot,5), (5,6), (6,7)\}$ . 
\begin{equation}
    \label{eq:alignment}
    \begin{matrix}
    dom(\alpha)& \texttt{1234~~56}\\
    q= & \texttt{TAGT\textemdash CA}\\
    \alpha= & \texttt{||~~~~||}\\
    t= &\texttt{TA-CGGCA} \\
    img(\alpha) & \texttt{12~34567}
    \end{matrix}
\end{equation}
The first two characters and the last two characters are identical, the fourth is a mismatch, and there are three gaps.
The query coverage of $\alpha$ is $QC(\alpha)=4/6$, the percent identity is $PI(\alpha)=\frac{2\cdot 4}{6+7}$, and the fraction of gaps is $Gaps(\alpha)=3/8$.
\end{example}

\subsection{Related work on DNA screening} 
\label{sec:screening}
\label{sec:hhsprotocol}
Some DNA sequences may encode extremely dangerous products, such as toxic peptides.
The Screening Framework Guidance for Providers of Synthetic Double-Stranded DNA (\textbf{HHS guidelines}), published by the United States Department of Health and Human Services, suggests methods to minimize the risk of unauthorized distribution of select agents~\cite{hhsscreening}. 
Highly related to the HHS guidelines are the Harmonized Screening Protocol v2.0 (HSP), employed by the International Gene Synthesis Consortium (IGSC)~\cite{dnascreening},
and the International Association Synthetic Biology (IASB) Code of Conduct for Best Practices in Gene Synthesis~\cite{IASBcode}.

The HHS guidelines, IASB Code of Conduct, and HSP outline standards and practices aimed at preventing the misuse of synthetic genes. 
They define procedures for customer screening and synthetic gene order screening for the presence of possible toxins, pathogens, and other biological agents that pose a significant threat to public health and safety, which are collectively referred to as \textbf{sequences of concern (\sat)}. 
US regulation also defines items on the Commerce Control List as sequences of concern.

While the HHS guidelines recommend screening specifically for the presence of sequences unique to sequences of concern, the HSP recommends identifying sequences derived from or encoding a sequence of concern. 
It is generally advised to use a sequence alignment tool, such as BLAST (the Basic Local Alignment Search Tool)~\cite{altschul1990basic}, to compare gene orders with known sequences in the GenBank database~\cite{benson2012genbank}. 
HHS guidelines recommend the \textbf{\bestmatch} approach to determine the legitimacy of an order based on the classification of the most similar sequence in the database. 
Specifically, every fragment of \textbf{200bp} in the order is searched within the database using sequence alignment. 
If the \bestmatch~of any fragment is a sequence of concern, the order is deemed a \textbf{hit}, and it is forwarded for further investigation. 
False hit rate is a major issue during screening. 

\textbf{GenoTHREAT}~\cite{adam2011strengths} software implements the HHS guidelines. 
The query sequence is partitioned into 200bp fragments. 
We evaluate the disclosed attack and defence against \genothreat and our own rigorous implementation of the HHS Guidelines based on \blast. 
Another software relying on sequence alignment for order screening is BlackWatch~\cite{blackwatch2005}. 
Early versions of both \genothreat~and BlackWatch were released as open source but were retracted later due to biosecurity concerns.
Note that general security practice advises against security by obscurity, striving instead for rapid identification and elimination of security weaknesses by the public.   
A recently published \textbf{NNTox}~\cite{jain2019nntox} screening tool employs Gene Ontology (GO) Terms and machine learning for predicting the toxicity of a sequence.    
Unfortunately, the published NNTox models failed to identify conotoxins and other short toxic peptides during preliminary tests and thus was not included in the evaluation.

In addition to biosecurity tools, there are multiple systems such as InterProScan\cite{jones2014interproscan}, SeqScreen\cite{albin2019seqscreen}, and various machine learning approaches~\cite{kulmanov2018deepgo,zhang2019deepfunc,elworth2020synthetic} that can be used to predict the function of DNA and protein sequences. 
These approaches provide detailed information for the human analyst to investigate a hit including both suspicious and legitimate elements  found within the DNA. 
In this article, we primarily argue against the \bestmatch~principle in DNA screening which allows an adversary to evade detection. 
Similar to the cyber warfare, the penetration of machine learning and adversarial learning into biosecurity domain should increase.

\vspace{-0.2cm}
\subsection{HHS guidelines criticism and responses}
\vspace{-0.2cm}
\label{sec:responses}
There are multiple concerns regarding the HHS guidelines~\cite{igsc}, including the possibility of assembling a \sat~using new bioengineering tools.
For example, using Gibson assembly, \textbf{oligos} (short DNA molecules) may be assembled to construct larger fragments~\cite{gibson2009enzymatic}. 
Since the HHS guidelines do not address oligonucleotide screening, one might order small DNA fragments which are not screened~\cite{kobokovich2019genesynthesissecurity} and can be assembled to create pathogens.

The HHS guidelines do not specify a database to use for screening, but it does suggest the GenBank as an example of such a database.
The lack of a formal database of \sat s~is a problem, since it may lead to inconsistent screening protocols between companies, false positives due to housekeeping genes shared between pathogenic and non-pathogenic organisms, and increased cost of overall screening~\cite{kobokovich2019genesynthesissecurity}. 
Housekeeping genes are required for basic cellular functions but do not produce toxins or other dangerous products.  
Nevertheless, housekeeping genes have long been known to cause false hits~\cite{bernauer2008technical}. 
A curated database of \sat s~may reduce the cost of sequence screening~\cite{carter2015dna,albin2019seqscreen}.
Other concerns are the possibility of poisoning public  databases with adversarial sequences that have been misclassified as benign, whether mistakenly or with malicious intent~\cite{caswell2019defending}.

The HHS guidelines recommend screening all sequences ordered, regardless of their length. 
Some argue that screening DNA fragments that are shorter than 200bp may lead to false positives and increased cost \cite{adam2011strengths}, while others argue that the 200bp cutoff is not scientifically justifiable \cite{tucker2010double}. We provide justification for a 60bp cutoff in Section~\ref{sec:seqobfuscation}.

\vspace{-0.2cm}
\subsection{Related cyber attacks}
\vspace{-0.2cm}
Cyberbiological attacks are reminiscent of attacks on physical assets, such as critical infrastructure or industrial control systems, from  cyberspace~\cite{mo2011cyber,kravchik2018detecting}. 
They are perhaps most similar to attacks on additive manufacturing~\cite{yampolskiy2018security}, where malware on the victim's computer manipulates blueprints of parts being produced on a 3D printer. 
Those parts may function (or malfunction) far away from both the cybercriminal and the victim, extending the reach of the former. 
Similarly, here the adversary affects a biological substance without direct contact with it. 
The fundamental difference between cyberbiological attacks and cyberphysical attacks is that, DNA can and should be considered as an executable code, although, executable in a non-digital environment.
A defect inserted by a malware into a 3D printed rotor may cause a drone to crash. 
But malicious gene injected into a DNA order by a malicious browser plugin may potentially subvert a DNA sequencing machine propagating back to the cyberspace. 
This potential bio-cyber attack presented by Ney et al. in USENIX Security'17~\cite{ney2017computer} closes the loop with the cyber-bio attack discussed here.

\vspace{-0.2cm}
\section{The biosecurity scenario}
\vspace{-0.2cm}
\label{sec:biosecurity}
Consider a case where a bioterrorist in California wants to build a synthetic virus or produce dangerous toxins, and distribute them locally. 
Synthetic DNA providers, members of IGSC, are the first line of defense. 
Following is a list of working assumptions considered in this section regarding the attacker and the defender:

\noindent(1) The attacker orders long DNA sequences. We consider oligos out of scope for this paper.  
\\\noindent(2) The attacker places an order with one of the synthetic DNA providers that screen orders. There are no other assumptions regarding the attacker.   
\\\noindent(3) The defender automatically screens every order with a sliding window of fixed size (200bp but may change) and hits are determined according to the \bestmatch~approach. 
This assumption is well aligned with the HHS guidelines. 
\\\noindent(4) Only when there is a hit during automatic screening is the order forwarded for further screening by a human analyst or by using DNA function prediction tools.   
This assumption stems from the large number of orders, the large size of DNA databases used for screening, and the significant computational resources that are therefore needed~\cite{igsc,elworth2020synthetic}. 
\\\noindent(5) We assume that all the malicious DNA sequences that the attacker tries to order are found in the defender's database. 
It is a worst-case assumption that makes the obfuscation process described next more challenging.     
    
\begin{algorithm}[t]
\small
\DontPrintSemicolon
\KwInput{\\
$\sat$ -- a Sequence of Concern\;
\screen{} -- a black-box screening algorithm\;}
\KwOutput{$O$ obfuscated Sequence of Concern}
\tcp{
@post-condition: Expressing $O$ in \cas~containing environment results in an assembly of $P$}

Partition \sat~into 64bp fragments $\sat=f_1+f_2+\ldots+f_n$\;
$f_1\leftarrow$ promoter and ribosome binding site $+f_1$\;
$f_n\leftarrow f_n+$ terminator\;
Let $m$ be a 23bp long efficient \cas~cutting site~\cite{cui2018review}\;
$m_{RC} \leftarrow$ reverse complement of $m$\;
$\forall_{1\leq i<n},f_i\leftarrow f_i+m$\;
$\forall_{1< i\leq n},f_i\leftarrow m_{RC}+f_i$\;
$Body = \sum_{i=1}^{n} $\hide{$f_i$}\;
$\forall_{1\leq i<n},hdr_i\leftarrow$32bp suffix of $f_i+$ 32bp prefix of $f_{i+1}$\;
Let $grs$ be a gRNA scaffold targeting $m$\; 
\tcp{$grs$ also targets $m_{RC}$ on the opposite strand.}
$Decoder = grs+\sum_{i=1}^{n-1} $\hide{$hdr_i$}\;
\tcp{Assemble and return the obfuscated sequence}
\KwRet{$O=Body + Decoder$}\;

\Fn{\hide{$x$}}{
    \tcp{Find benign gene $c$ that is most similar to $x$.}
    Find best scored $(c,\alpha) \in$ \blast{$x$,no~gaps} such that $img(\alpha)$ is at least 200pb from $c$'s ends\; 
    $cx\leftarrow$ replace $\alpha(x)$ in $c$ with $x$\; 
    \eIf{\screen{$cx$}=hit}{
        \KwSty{error} obfuscation failed
    }
    {
        \KwRet{cx}\;
    }
  }
\caption{\sat~Obfuscation 2 (\osatwo)\label{alg:osa2}}
\vspace{-0.1cm}
\end{algorithm}

\vspace{-0.2cm}
\subsection{Below the DNA screening radar}
\label{sec:attack}
\label{sec:seqobfuscation}

At a glance, a sequence of concern (\sat) that needs to be obfuscated is split into small fragments. 
The fragments are interleaved with legitimate DNA sequences which are as similar as possible to the \sat~fragments. 
As a result, for every 200bp window the \bestmatch~is always a legitimate sequence.  
This may sound like a trivial obfuscation against pattern-based detectors that long ago became obsolete in cyberwarfare, but its biological implementation poses several challenges, as discussed below.

We note that \genothreat~can easily be evaded due to its strict conditions for identifying a hit.
We omit the details 
here\footnote{See Appendix~\ref{sec:osa1} for details on \osaone.} because it is an attack against a specific vulnerable algorithm. 
Later the specific attack against \genothreat~is referred to as level 1 \sat~obfuscation (\osaone) and the generic attack described next as level 2 \sat~obfuscation (\osatwo).
%
%
%

\begin{figure*}[t]
    \centering
    \includegraphics[width=0.95\textwidth]{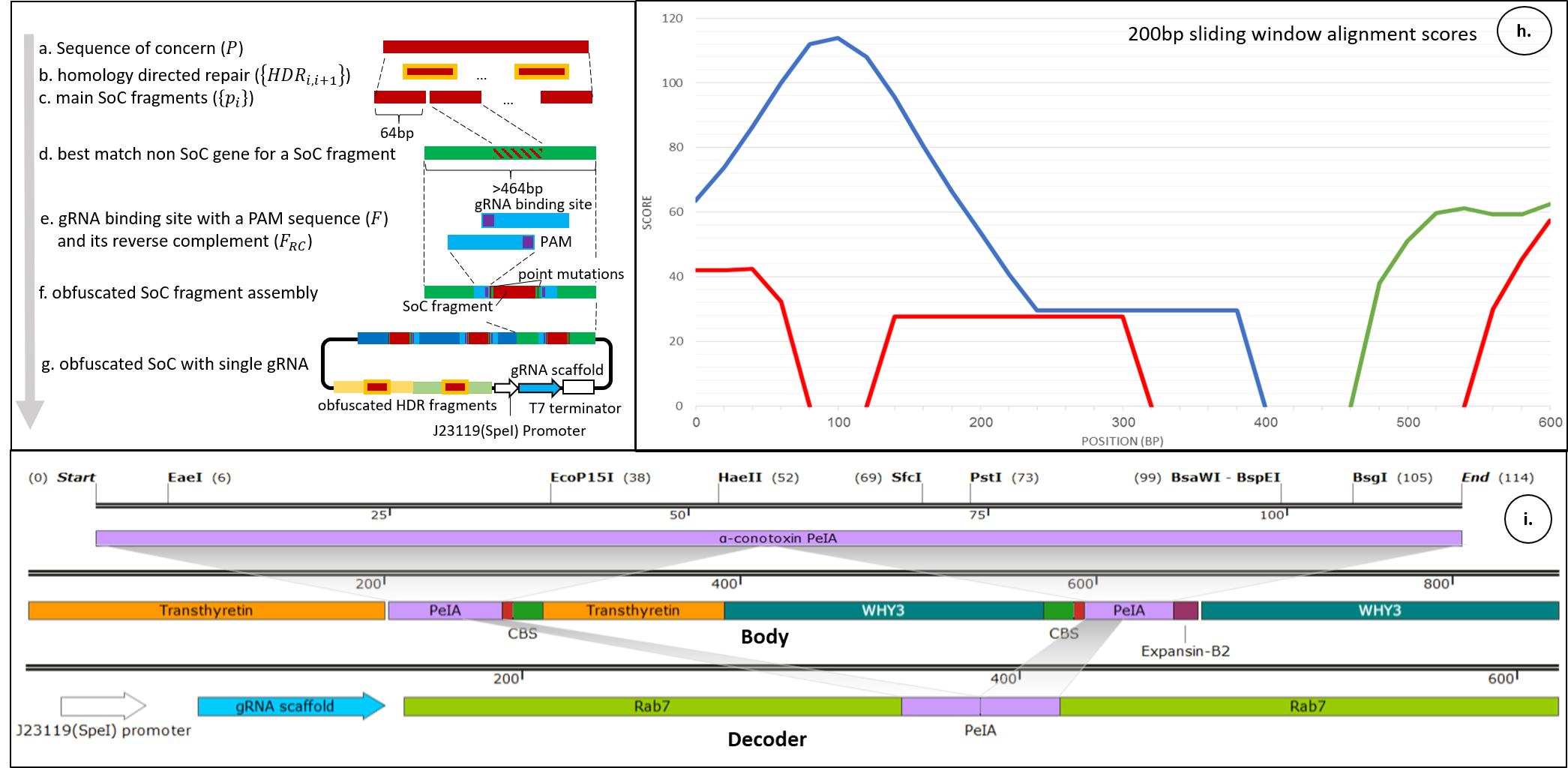}
\caption{Level 2 \sat~obfuscation. Schematics of the \sat~obfuscation process (a-f); a DNA containing obfuscated $\alpha$-conotoxin PeAI and a decoder sequence that facilitates its reconstruction (g); \texttt{blast} scores when scanning an obfuscated sequence with a 200bp sliding window (h). Colors correspond to the schematic on the left: green and blue lines are alignment scores with camouflage genes, and the red line is the alignment score with the \sat~subject.}
    \label{fig:osa2}
    \vspace{-0.4cm}
\end{figure*}

\label{sec:osa2}
Level 2 obfuscation allows a \sat~to remain below the screening radar of protocols implemented according to the HHS guidelines.  
Specifically, we exploit the specification stating that if for every 200bp fragment of a query sequence the \bestmatch~is \nsa, then the query is not a hit.
The general obfuscation process is depicted in Figure~\ref{fig:osa2}a-f and Algorithm~\ref{alg:osa2} presents its pseudocode. 
We will refer to the figure and the pseudocode while presenting \osatwo.

The main challenges an attacker faces in order to successfully evade detection are: 
\\\noindent(1) Finding legitimate DNA that can successfully camouflage the \sat~fragments. 
\\\noindent(2) Decoding the obfuscated DNA with a minimal set of standard biological primitives commonly used in experiments.

\paragraph{The camouflage genes}
Given a \sat, we split it into small fragments ($SoC=f_1+\ldots+f_n$).
Plus $+$ denotes string concatenation. 
We will refer to the length of these fragments later in this subsection.
In order to hide a \sat~fragment we search for a \nsa~gene that is most similar to the fragment. 
We will refer to such \nsa~genes as \emph{camouflage genes}.

The search for camouflage genes can be performed within standard nucleotide and protein databases using \blast~\footnote{\url{https://blast.ncbi.nlm.nih.gov/Blast.cgi}}.
In order to efficiently find camouflage genes we search for each \sat~fragment ($f_i$) while excluding organisms that are known to contain the \sat~or similar compounds.   
The search is performed with the lowest mismatch penalties and the highest gap penalties. 

In the following discussion, let $c$ denote the appropriate camouflage gene and $\alpha$ denote an alignment between $c$ and $f_i$.      
In order to hide the \sat~fragment within the camouflage gene we replace the respective nucleotide sub-sequence ($img(\alpha)$) with the \sat~fragment $f_i$. 
This process is summarized in Algorithm~\ref{alg:osa2} function \hide. 
See also Figure~\ref{fig:osa2}~a,~c,~and~d. 


The \sat~fragments embedded within the camouflage genes constitute the main body of the obfuscated DNA. 
In order to properly function within living cells after decoding, the body also needs a promoter, a terminator, and a ribosom binding site. 
See lines 1-3, 8 in Algorithm~\ref{alg:osa2}. 
The target organism chosen can be one of the most prevalent organisms used in DIY biology, for example, E. Coli.

\paragraph{The decoder}
Similar to the old well-known encrypted or oligomorphic malware~\cite{rad2012camouflage}, building a decoder is the most challenging part in designing obfuscated malicious DNA.
Here the decoder should operate inside living cells rather than in cyberspace. 
Decoding is facilitated by the maturity and prevalence of CRISPR-based DNA editing methods. 
In general, the decoder needs to perform two tasks when reconstructing the malicious DNA: \
(1) It should remove the parts of the camouflage genes between consecutive \sat~fragments. 
(2) It should stitch consecutive \sat~fragments together, forming operational DNA.  

In order to cut out parts of the camouflage genes during decoding, we add cut points before and after each \sat~fragment ($f_i\text{\ding{34}}camouflage\text{\ding{34}}f_{i+1}$). 
Cut points are marked using a 23bp sequence $m$ which includes PAM and gRNA target site.  
The following example of such a sequence is cut at the nucleotide marked with \ding{34}. 
\begin{equation}
\label{eq:F}
m=\scriptsize \begin{matrix}
\texttt{CCTTCC} & \texttt{ACAAGCTCGCCGAGGTG}\\
\texttt{GGAAGG} & \texttt{TGCTCGAGCGGCTCCAC}\\
\underbrace{\texttt{~~\ding{34}~~~}} & \underbrace{\texttt{~~~~~~~~~~~~~~~~~}} \\
\text{\scriptsize PAM} & \text{gRNA target sequence}  
\end{matrix}
\end{equation}
It is important to note that CRISPR may bind to either side of the double-strand DNA molecule to perform the cut. 
Therefore, a reverse complement of $m$ (denoted $m_{RC}$) is cut from the opposite strand, as shown in Equation~\ref{eq:FRC}:
\begin{equation}
\label{eq:FRC}
m_{RC}=\scriptsize \begin{matrix}
\text{gRNA target sequence} & \text{ PAM} \\
\overbrace{\texttt{~~~~~~~~~~~~~~~~~}} & \overbrace{\texttt{~~~\ding{34}~~}} \\
\texttt{CACCTCGGCGAGCTCGT}& \texttt{GGAAGG}\\
\texttt{GTGGAGCCGCTCGAGCA}& \texttt{CCTTCC}
\end{matrix}
\end{equation}

The gRNA target sites within the main body point to the locations that need to be cut (see Algorithm\ref{alg:osa2} lines 6-7 and Figure~\ref{fig:osa2}~e~and~f). 
In order to actually perform the cuts the decoder should include the respective gRNA scaffold targeting the cut points (see Algorithm\ref{alg:osa2} line 10 and Figure~\ref{fig:osa2}~g).
Cas9 that is also required to perform the cuts is a part of standard gene editing instrumentation which can easily be purchased online (Figure~\ref{fig:diybio}).  

Next, there is a need to repair the DNA cut made by CRISPR when removing the camouflage and to repair the residue PAM nucleotides\footnote{For example, the leftmost \texttt{GG} on the lower strand in Equation~\ref{eq:F}.}. 
Such a repair can be performed using the HDR process which requires a 64bp long template ($hdr_i$) -- a DNA sequence covering both the last 32 nucleotides of $f_i$ and the first 32 nucleotides of $f_{i+1}$ (see Algorithm~\ref{alg:osa2} line 9 and Figure~\ref{fig:osa2}~b). 
A 64bp long HDR template is the shortest HDR template shown to perform well in practice~\cite{pohl2016crispr}. 
The longer the HDR templates are, and the more replicas of these templates are found within the cell, the higher the HDR efficiency is.
Since $\{hdr_i\}$ are also fragments of the \sat~sequence, it is important to camouflage them in order to remain below the screening radar.

\paragraph{Obfuscation variants} 
There could be many variants of this general obfuscation scheme, including variations of the gRNA binding sites.  
On the one hand, these sites may be designed individually for each cut point in order to improve the stealthiness of the obfuscation. 
But in this case they will require multiple gRNA scaffolds, each targeting a different cut point, potentially reducing the decoding's effectiveness. 
On the other hand, repeated gRNA binding sites may increase the likelihood of obfuscation failure.

The size of the \sat~fragments and the HDR templates may also vary. 
Reducing the size of the \sat~fragments increases their likelihood of blending within the camouflage genes.  
Yet it also increases the number of such fragments, which leads to a larger number of cuts and repairs, consequently reducing the effectiveness of the decoder.  
Since 64bp is the currently known lower bound on the size of HDR templates, there is no point in using shorter \sat~fragments if they are ordered together with the HDR templates. 
The adversary may also choose to order the main body and the HDR templates separately and reduce the size of the \sat~fragments.

\vspace{-0.2cm}
\subsection{\ged}
\vspace{-0.2cm}
\label{sec:ged}
In order to harden synthetic DNA order screening and reduce the non-regulated distribution of select agents and toxins, we propose a new DNA screening algorithm termed \ged~(GED).  
The algorithm is designed to assess the difficulty of \sat~assembly from a DNA sequence. 
In order to do so, GED screens the query sequence to find all substrings which are similar to fragments of a \sat. 
Then, GED quantifies the effort of assembling a \sat~from these fragments.
Although designed with a focus on \sat, GED can quantify the effort required to assemble any target sequence $t$ from a query sequence $q$ using a standard CRISPR system.
More specifically, we count the number of cuts and repairs required for constructing the target sequence from the query sequence. 

In a standard biological sequence alignment, a typical objective is to identify genes conserved in different genomes.  
To achieve this objective, the match reward ($rm$), mismatch penalty ($pmm$), and gap penalties ($gpo$, $gpx$) must be well balanced when computing the alignment score. 
GED's objective is more complex. 
On the one hand, we need to find short conserved regions with minimal gaps within the query sequence. 
On the other hand, we want to concatenate the short conserved regions regardless of the gap length between them.

\begin{example}
\label{ex:peia-blast}
A default configuration when aligning two sequences with \blast~\cite{altschul1997gapped} is $rm=2$, $pmm=3$, $pgo=5$, and $pgx=2$.
Figure~\ref{fig:peia-blast} presents the results of aligning the obfuscated $\alpha$-conotoxin PeIA\footnote{A short toxic peptide.} from Figure~\ref{fig:osa2} to the PeIA sequence using default configuration. 
\blast returns three ranges (alignments). 
As long as $pgx>0$, these ranges will not be merged by \blast due to the length of the gap that would be opened in the target sequence. 
However, when removing a sequence between two consecutive \sat~fragments using the CRISPR system and HDR templates, the distance between them does not play a critical role.  
\end{example}

\begin{figure}
    \centering
    \includegraphics[width=0.5\columnwidth]{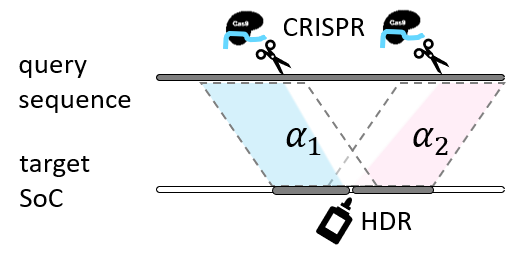}
    \vspace{-0.2cm}
    \caption{Two local alignments with disjoint domains and overlapping images.}
    \label{fig:merge}
    \vspace{-0.4cm}
\end{figure}

\paragraph{Gap removal penalty}
\label{sec:prm}
In order to achieve the objective of GED, we introduce a new gap removal penalty ($prm$) that substitutes some gap opening and extension penalties within the target sequence.\footnote{Note that variation of gap penalties for query and target sequences has successfully been used in the past for other use cases~\cite{thompson1995introducing}.} 
Let $g=[a,b]$ be a gap in the target sequence ($\forall_{i\in g}, \alpha(i)=\bot$). 
Removal of $[a,b]$ from the query sequence requires two cuts, at $a$ and at $b$ respectively, followed by an HDR (e.g., Figures~\ref{fig:merge}~and~\ref{fig:attack}.10).   
These operations may fail. 
Let $\gamma$ be the probability that a gap is successfully removed.  

To define GED we neglect some biological constraints of the CRISPR system, such as specific PAM sites targeted by \cas, assuming that cuts can occur at any point in a DNA sequence.
This is a worst-case assumption overestimating the potential risk. 
Future versions of GED may take biological constraints into account to reduce false positives. 
Currently, in the absence of advanced adversarial techniques, GED detects obfuscated \sat~with 100\% accuracy, as shown in Section~\ref{sec:eval}.

If the gap $g=[a,b]$ is removed, the score of the alignment will increase by $pgo+pgx\cdot(b-a+1)$ and reduce by $prm$. 
Let $\mathcal{G}_{q}(\alpha)$ and $\mathcal{G}_{t}(\alpha)$ be sets of all gaps in the query and target sequences respectively according to the alignment $\alpha$. 
Let $\mathcal{G}^k_t\subseteq \mathcal{G}_t$ be a subset of the $k$ longest gaps in the target sequence to be removed. 
The probability that all gaps in $\mathcal{G}^k_t$ will be successfully removed is $\gamma^{k}$. 
Next, we adjust the alignment score as a result of designating $\mathcal{G}^k_t$ for removal.  
\begin{definition}[Adjusted alignment score]
\label{def:gap-removal}
Given an alignment $\alpha$; a subset of gaps $\mathcal{G}^k_t$ in the target sequence of $\alpha$; gap opening and gap extension penalties $pgo$ and $pgx$ respectively; gap removal penalty $prm$; and gap removal probability $\gamma$; we define the change in the alignment score as the result of the removal of the $k$ longest gaps in the target sequence $t$ as:
$$
Score^k(\alpha)= Score + \gamma^{|\mathcal{G}^k_t|}\cdot\sum_{g\in\mathcal{G}^k_t} \left(pgo+pgx\cdot|g|-prm\right).
$$
\end{definition}

Next, we examine the choice of gaps to be removed ($\mathcal{G}_{trm}$) and the parametrization of $Score^k$. 
We assume that the match reward ($rm$), as well as the mismatch, gap opening, and gap extension penalties ($pmm,pgo,pgx$) are set at their default values or according to the biological considerations (which are beyond the scope of this article).

$Score^k$ approaches $Score$ when the number of gaps to be removed $k$ increases and $\gamma<1$. 
The choice of $\gamma$ depends on the assumptions made regarding the expected biological effectiveness of the attacks given typical bioengineering tools used by potential victims today.    
$\gamma=0$ signifies that no attacker could ever rely on residue \cas~protein in the cells and the \sat~obfuscation attack described in Section~\ref{sec:seqobfuscation} is impossible.  
$\gamma=1$ signifies 100\% success of gene editing, which leads to the successful decoding of \osatwo~sequences regardless of the number of \sat~fragments.  

The gap removal penalty $prm$ should be set to a value that would justify gap removal using bioengineering tools. 
For example, if only the removal of gaps larger than $x$bp is justified, then the gap removal penalty should be set to:
$$
prm = pgo + pgx\cdot x.
$$
Obviously, the longer a gap is, the more worthwhile its removal is.  
Thus, for the sake of computing $Score^k$, only the longest gaps are selected.
The number of gaps $k$ that should be included in $\mathcal{G}^k_t$ for the maximal $Score^k$ depends on the parameters and can be selected using a grid search or simple hill climbing algorithm.

\begin{figure}
    \hspace{-0.4cm}\includegraphics[width=1.10\columnwidth]{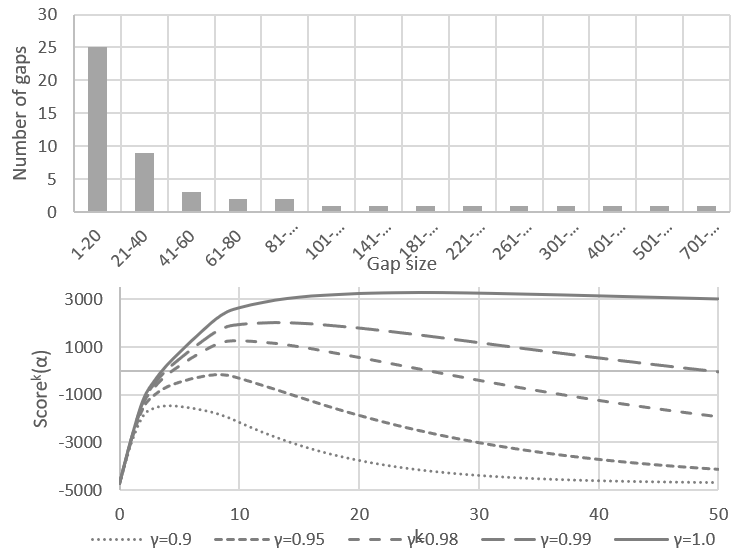}
    \caption{\label{fig:ex-dscore}
    Example of an alignment score correction using the GED $\Delta Score$. (a) The gap length distribution of an example alignment. (b) The corrected alignment score ($Score+ \Delta Score^k$) for various choices of $k$.}
    \vspace{-0.3cm}
\end{figure}

\begin{example}
Assume, for example, that we examine a 10Kpb long query sequence, comparing it to a 2Kpb long \sat. 
Also assume a negatively scored optimal alignment between the two sequences which contains 2K matching base pairs, a few mismatches, and 50 gaps whose length is exponentially distributed, as depicted in Figure~\ref{fig:ex-dscore}a.  
Although, such alignment exists, it will never be returned by the \blast search engine, because its score is worse than random. 
Assuming default \blast configuration, the specific alignment score could be lower than $-4,500$. 
Assume a gap removal penalty of $prm=20$.  
Replacing gap opening and gap extension penalties with $prm$ starting from the longest gaps would increase the score, as shown in Figure~\ref{fig:ex-dscore}b.    
With $\gamma=0.98$, setting $k=10$ results in the highest $Score^k$ value. 
With $\gamma=0.99$, the adjusted alignment score can reach $2,000$ when $k=13$ in this example.  
\end{example}

\begin{example}
Further following Example~\ref{ex:peia-blast}, consider the alignment of obfuscated $\alpha$-conotoxin PeIA with the unobfuscated peptide, as shown in Figure~\ref{fig:peia-blast}. 
The blue alignment in Figure~\ref{fig:peia-blast}c, was not identified by \blast, because it contains very long gaps in the target sequence. 
Nevertheless, an adjusted score with the gap removal penalty of $prm=20$ and $\gamma=0.99$, according to definition~\ref{def:gap-removal}, would result in a total score of 193 which is very close to a best match. 
\end{example}

\paragraph{GED definition}
Next, we define the gene edit distance as the optimal number of gaps ($k$) to remove in a query sequence 
such that the adjusted alignment score is maximized.
\begin{definition}[Unidirectional gene edit distance]
The gene edit distance (GED) from $q$ to $t$ is:  
$$
GED(q,t) = \texttt{ARGMAX}_k \texttt{MAX}_{\alpha} Score^k(\alpha).
$$
\end{definition}
According to this definition $Score^{GED(q,t)}(\alpha)$ has the maximal adjusted score.
Although we call GED a distance, it is not a valid mathematical distance metric, first and foremost because it is asymmetric. 
GED quantifies the effort required to transform $q$ into $t$ but not vice versa. 
As an academic exercise, one can define a true gene edit metric, but such definition is beyond the scope of this security article.

\paragraph{GED computation} 
Current \blast implementations will not return suitable alignments which minimize the small gaps but neglect the long ones.  
Thus, in order to reduce time-to-market and maintain backward compatibility with existing \blast engines, we implement a GED heuristic as postprocessing of standard \blast outputs.   
The GED heuristic pseudocode is presented in Algorithm~\ref{alg:ged}.

\begin{algorithm}[t] \label{alg:ged}
\small
\DontPrintSemicolon
\KwInput{\\
q -- a query sequence\;
t -- a target sequence\;
}
\KwOutput{ \\
$k$ -- the number of cut and repair operations\;
$Score^k$ -- adjusted alignment score\;
}
\tcp{List local alignments $\mathcal{A}=\{\alpha\}$ sorted by $dom(\alpha).start$}
$\mathcal{A}_{q,t} = \blast{query=q,subject=t}$ \;
\For{each subset $P\subseteq \mathcal{A}_{q,t}$ of local alignments with disjoint domains}{
    $\alpha_P = \merge{P}$\;    
}
$P^*=\texttt{argmax}_{P} Score^{|P|-1}(\alpha_{P})$\;
\KwRet{$k=|P^*|, Score^k(\alpha_{P^*})$}\;
\Fn{\merge{$\alpha_1,\ldots,\alpha_k$}}{
    \tcc{Find global alignment between unified domains and $t$}
    $\mathcal{A}=\blast(query=q[\bigcup_i dom(\alpha_i)],subject=t)$\;
    $\alpha=$ cleaned global alignment unifying $\mathcal{A}$\;
    \tcc{Reintroduce gaps to form a continuous alignment}
    $\alpha\cup=\{(i,\bot) : i\in[dom(\alpha_j).end,dom(\alpha_{j+1}).start], 1\leq j<k \}$
    \KwRet{$\alpha$}
}
\caption{Gene Edit Distance}
\end{algorithm}

Standard \blast algorithms compute a large set of small local alignments and extend these alignments when doing so increases the alignment score.
We take a similar approach when computing a set of alignments using standard \blast configuration and merging them to maximize $Score^k$. 
Recall that $\mathcal{A}_{q,t}$ is a set of local alignments between $q$ and $t$ returned by \blast (Algorithm~\ref{alg:ged}, line 1).
Let $.start$ and $.end$ denote the first and the last position respectively, in the domain or the image of an alignment $\alpha\in \mathcal{A}_{q,t}$.   
Let $P=\alpha_1,\ldots,\alpha_k$ be a set of alignments whose domains are disjoint, $dom(\alpha_i).end<dom(\alpha_{i+1}).start$ (Algorithm~\ref{alg:ged}, line 1).
Their images may overlap, as depicted in Figure~\ref{fig:merge}.   
\blast can find a global alignment between the unified domains of the alignments and the target sequence i.e. when sequence fragments between the alignment domains are removed.   
Such unified alignments are called cleaned alignments according to \blast.

The output of Algorithm~\ref{alg:ged} is the number of cut and repair actions required to reconstruct the target sequence from the query sequence and the adjusted alignment score. 
The latter quantifies both the similarity of the body of an obfuscated \sat~and the effort required to decode it within the cells.
Unlike in the case of encrypted or oligomorphic malware where decoders are the easiest to detect, here we concentrate on detecting the main body of the obfuscated \sat, because the gRNA scaffolds and HDR templates comprising the decoder may be distributed among different plasmids or even different orders.


In order to return a final assessment of the risk a query sequence presents, the GED of the query sequence is evaluated against all \sat~sequences in the database. 
The final judgement is made based on the maximal adjusted score of the query sequence for any of the \sat~sequences. 
This allows the identification of seemingly benign sequences that can easily be transformed into malicious sequences that produce dangerous products. 
In addition, such an approach is more resilient to an attack in which a public gene database is poisoned with legitimate sequences, although poisoning with legitimate sequences maliciously marked as a \sat~may result in false hits and require human attention.

\vspace{-0.2cm}
\subsection{Evaluation}
\vspace{-0.2cm}
\label{sec:eval}

\label{sec:gtp}
In this section we evaluate different screening algorthms vs. the \osaone~and \osatwo obfuscation algorithms.

\paragraph{Benchmark dataset}
\label{sec:screencompare}
For the purpose of evaluation  
we selected 50 \sat s from the UniProt database~\cite{uniprot2019uniprot}, all of which are marked as toxins, are manually reviewed, and contain between 33 and 100 amino acids.
These protein sequences were reverse translated to sequences of nucleotides using an EMBOSS online tool~\cite{madeira2019embl}. 
We ran \blast queries using the nucleotide \sat s to ensure that they are well detected within the GenBank database using \blast. 
Doing so, we identified seven benign sequences with various levels of similarity to the \sat s.   
We selected an additional 43 benign sequences from UniProt, non of which are toxic and all of which are manually reviewed.
The 50 \sat~sequences were obfuscated using Algorithms~\ref{alg:osa1} and~\ref{alg:osa2}.
To create the \osatwo~sequences, we selected the camouflage genes from the 50 \nsa~sequences described above.  
Finally, we generated 50 random nucleotide sequences (denoted as Rnd) as a reference point.   
There are a total of 250 nucleotide sequences in the benchmark dataset, equally split among \nsa, \sat, \osaone, \osatwo, and Rnd.




\paragraph{Baseline screening algorithms}
\genothreat~can be easily circumvented, because it scans non-overlapping 200bp fragments of the query sequence and employs very strict match constraints (see  Appendix~\ref{sec:osa1} for details).    
In order to objectively assess the threat of DNA obfuscation, we implement a screening method, referred to as  GenoTHREAT plus (\gtp), that strictly implements HHS guidelines.
While we don't elaborate on the algorithm here due to space constraints, we provide some highlights below.  


\gtp~scans the query sequence with a sliding window of length 200bp and step size of 1bp. 
Every such 200bp nucleotide sequence is searched in the database.
We use a set of keywords and anti-keywords to identify dangerous substances in the returned results. 
\gtp~records the score of the best matching \sat~as well as the score of the best matching \nsa. 
Later, we derive a confidence value from the difference between them. 
Querying both nucleotide and protein databases for each sequence of 200bp is a highly inefficient, yet accurate approach that strictly follows the HHS guidelines.   

In addition to \genothreat~and \gtp, we employ a heuristic screener based on InterProScan 5~\cite{jones2014interproscan}, a state of the art DNA and protein functional annotation framework. 
The heuristic referred to as IPC, produces a hit if at least one of the tools within InterProScan annotates at least some part of the evaluated DNA as a toxin. 

\paragraph{Results}
Benchmark sequence screening was performed using local copies of the complete GenBank NT and NR and UniProt databases.  
While \genothreat, IPC, and \gtp~require the entire database for accurate screening, GED only requires the \sat s~to compare with the query sequences.

\begin{figure}
    \centering
    \includegraphics[width=\columnwidth]{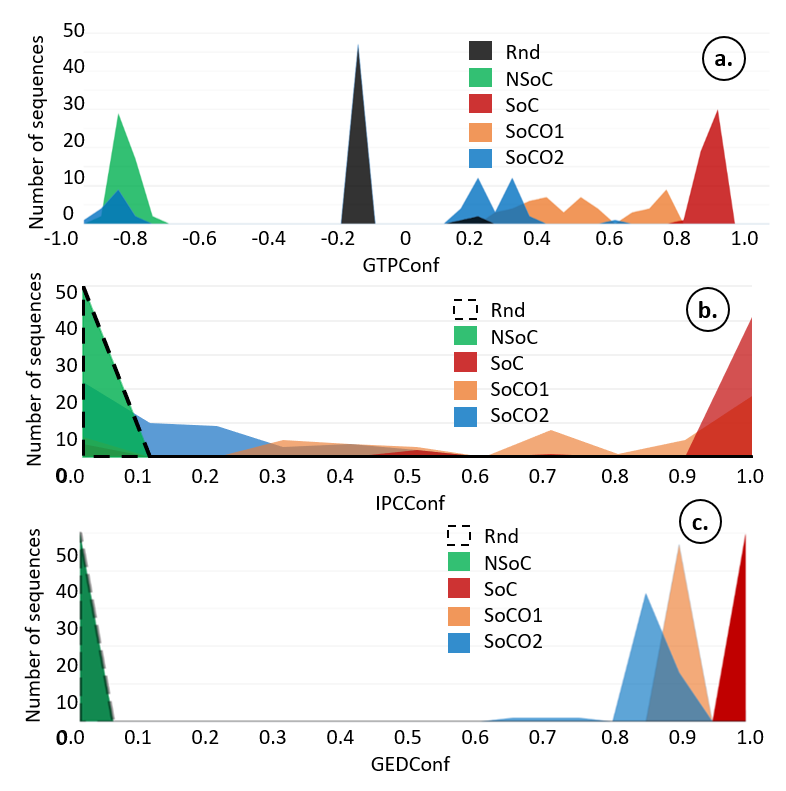}
    \vspace{-1.0cm}
    \caption{\gtp, IPC, and GED confidence levels for five types of sequences in the benchmark dataset.}
    \label{fig:confidence}
    \vspace{-0.5cm}
\end{figure}

\begin{table}[]
    \centering
    \caption{\label{tab:results}Number of hits produced by different screening approaches. }
    \begin{tabular}{|l|c|c|c|c|}
     \hline
                   & \genothreat  & \gtp & IPC & GED \\
     \hline
     \sat~hits     & 50           & 50   & 45** & 50 \\
     \osaone~hits  & 9            & 50   & 44 & 50 \\
     \osatwo~hits  & 7            & 34   & 28 & 50 \\
     \nsa~hits     & 0            & 0    & 0 & 0 \\
     Rnd~hits      & 0            & 3    & 0 & 0 \\
     \hline
     TPR           & 0.44         & 0.89  & 0.87 & 1.0 \\
     FPR           & 0.0          & 0.03* & 0 & 0.0 \\
     FNR           & 0.56         & 0.1 & 0.13& 0.0 \\
     \hline
    \end{tabular}
    \flushleft
    {\scriptsize\linespread{0.5}
    
    * Due to hits on random sequences.\\
    ** Five short toxins were not found within InterProScan databases. We disregard them while computing TPR, FPR, and FNR (i.e. assume 135 toxins instead of 150). 
    
    }
    \vspace{-0.5cm}
\end{table}

We expect a good screening algorithm to report a hit when screening sequences in the \sat, \osaone, or \osatwo. 
We expect a non-hit when screening sequences in the \nsa~or Rnd.
Similar to studies focusing on malware detection, we consider the true positive rate (TPR), the false negative rate (FNR), and the false positive rate (FPR). 
High FNR (low FPR) indicate that the screening method can be evaded using obfuscation.  
Since some malicious sequences are easier to detect than others, we the hit counts for each group of sequences.  
In order to analyze the performance of the screening algorithms, we also inspect their confidence levels.  

While \genothreat~only provides a binary decision on a query sequence, \gtp~can provide a confidence level along with the decision. 
We compute the \gtp~confidence as follows. 
Let $q$ be some 200bp fragment of the screened sequence. 
Let $Max\sat (q)$ and $Max\nsa (q)$ be the highest score of an alignment of $q$ with a \sat~and \nsa~respectively.
If \blast~did not return alignments with \sat~or \nsa, we set the respective score to zero. 
We define \gtp~confidence as
$$
GTPConf = \max_q \left\{\frac{Max\sat(q)-Max\nsa(q)}{\max\{Max\sat(q),Max\nsa(q)\}}\right\}
$$
$GTPConf>0$ means a hit, because there is at least one 200bp fragment that is more similar to a \sat~than to an \nsa.

The confident of IPC is the fraction of toxin annotations out of all successful annotations produced by InterProScan when screening $q$.   
$IPCConf>0$ means a hit.

The confidence of GED ($GEDConf$) is simply the maximal adjusted score it returns when screening $q$.
The value of $GEDConf$ is between zero and one, where
$GEDConf=1$ means that $q$ is definitely a \sat.

Table~\ref{tab:results} summarizes the performance of \genothreat, \gtp, and GED on the benchmark dataset. 
Confidence levels of \gtp~and GED when screening the 250 benchmark sequences are presented in Figure~\ref{fig:confidence}.

As expected, all \sat s are the closest to being malicious (right part of the chart) and all \nsa s are the closest to being benign (left part of the chart).

Most random sequences are classified by \gtp~as benign (low negative confidence), because there are many more benign sequences than toxic sequences in the GenBank database.  
Nevertheless, some random sequences contain 200bp fragments that are more similar to a \sat~than to an \nsa.
One such fragment is sufficient to produce a hit and require human attention. 
We consider such statistical errors as an inherent deficiency of the \bestmatch~approach. 
According to GED, random sequences are similar to benign sequences with respect to the effort required to transform them into something dangerous.

Next, we observe a high variance of the confidence levels of obfuscated sequences in Figure~\ref{fig:confidence} for all screening methods.  
While \osaone~sequences are always detected, some  \osatwo~sequences remain below the radar of the \bestmatch~approach, represented here by \gtp adn below the radar of InterProScan functional annotation.   
This is due, of course, to the camouflage genes which contribute the most to the PI value in each 200bp fragment. 
The partitioning into fragments makes it harder for InterProScan to detect functional segments. 
Nevertheless, some \osatwo~sequences are detected, because no sufficiently similar camouflage gene was found during obfuscation. 
An important conclusion from these results is that some sequences are easier to obfuscate than others.  

GED successfully detects all obfuscated malicious sequences. Moreover the large confidence gap between the most suspicious benign sequence and the least suspicious malicious sequence ensures GED's robustness as a screening algorithm.  
We also note that GED is an order of magnitude faster than \genothreat,~because it compares the query sequence with a small database of \sat~and does not need to query for every 200bp fragment.

\vspace{-0.2cm}
\section{DNA injection as a cyberbiological attack}
\vspace{-0.2cm}
\label{sec:cyber-bio}

In this section we introduce an attack pattern called \emph{DNA Injection} following the common template of Attack Pattern Enumeration and Classification (CAPEC) descriptions.\footnote{\url{https://capec.mitre.org/}}   
In contrast to classical biosecurity scenarios, the attack pattern discussed here assumes that the attack originates from cyberspace as do all cyberphysical attacks.

\begin{figure*}[h!]
\centering
\includegraphics[width=0.9\textwidth, height=\textheight,keepaspectratio]{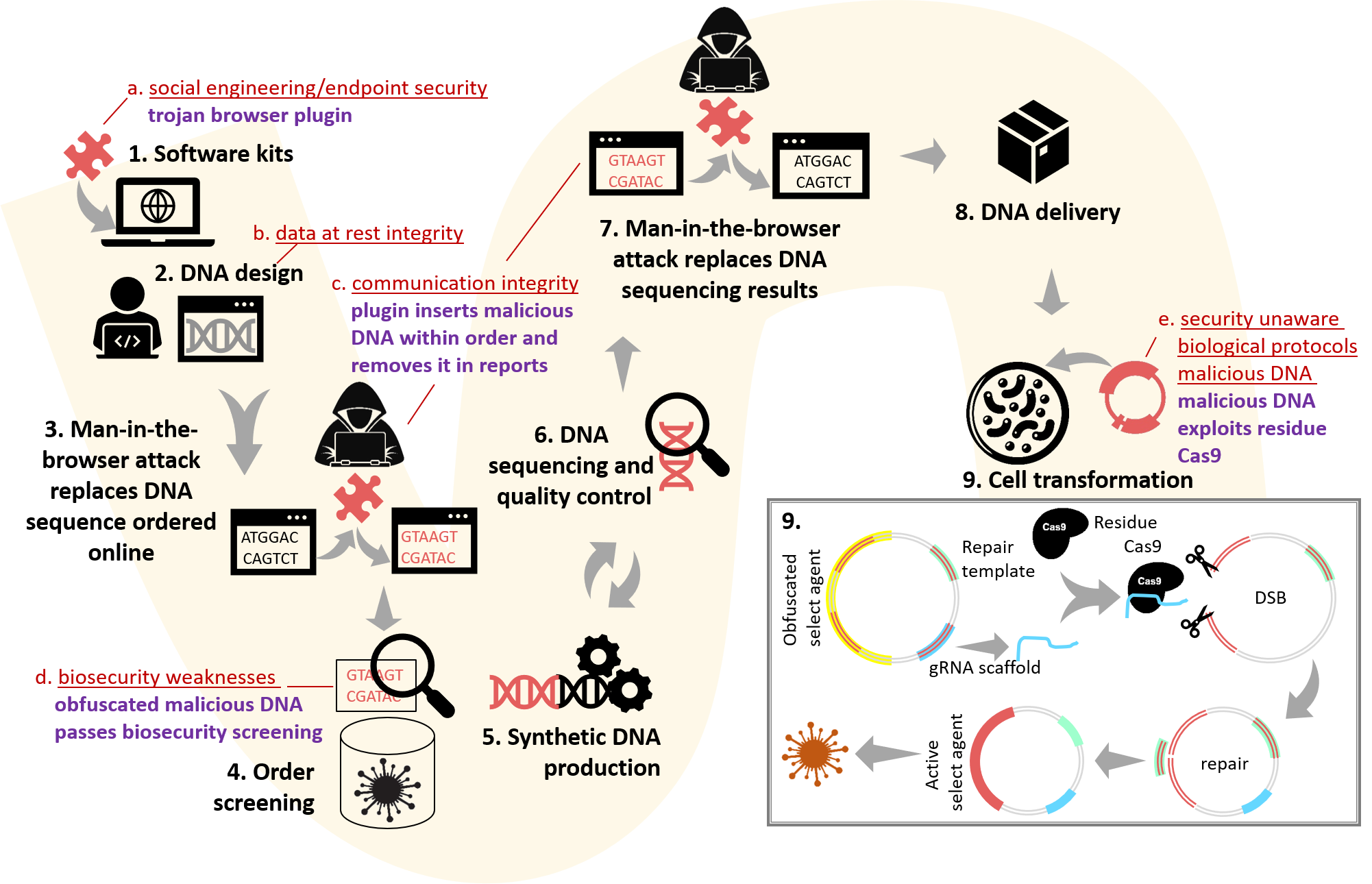}
\caption{\label{fig:attack}A typical synthetic biology workflow (1-9), weaknesses (a-e red underlined), and attack components (a,c,d,e purble bold). 
course of an \rsap~attack. The cell transformation schematic (9) depicts the decoding of the malicious DNA within a cell.
}
\vspace{-0.5cm}
\end{figure*}

In the discussed attack scenario, a \textbf{victim} might be a do it yourself (DIY) biology enthusiast or a small bioengineering company that develops their own DNA sequences or combines existing genes to produce fuel, medical components, or resilient plants. 
We assume that the victim does not use their own facilities to produce the DNA but prefers ordering synthetic DNA strands from synthetic gene \textbf{providers}.  
Figure~\ref{fig:attack} (steps 1-9) depicts a common workflow of a biologist experimenting with gene editing. 
First, the biologist designs a system for adding, removing, or replacing a gene in some target organism (e.g. E. Coli) using CRISPR, and then places a synthetic gene order with one of the providers, who screens the order for possible inclusion of \sat. 
During the production a provider typically sends the biologist updates on production status and quality control.
Once the DNA is produced, a tube is delivered to the biologist.  
Large labs prefer sequencing the order they received to ensure its quality, but DIY biologists and many small companies would trust the quality reports sent by the provider. 
Finally the synthetic genes are applied to the organism of interest and their biological effects are assessed.   
The \textbf{attacker} is a cybercriminal who wants to trick the biologist victim into producing dangerous biological components inside the victim's lab.  
For this purpose the attacker may design an obfuscated malicious gene as described in Section~\ref{sec:osa2}.

\vspace{-0.2cm}
\subsection{Related weaknesses}
\vspace{-0.2cm}
\label{sec:vuln}

\paragraph{Endpoint security}
DIY biologists are well trained in biosafety protocols and aware of biosecurity screening and the dangers of engineering pathogens, similar to most bioenginers~\cite{grushkin2013seven}.  
However, like most computer users, they cannot be trusted to properly secure their end devices and mitigate social engineering attacks. 
It is a common assumption that the end customer may be infected with malware (Figure~\ref{fig:attack}a) and providers as well as software vendors must be alert to that.

\paragraph{Data at rest} The first line of defense should have been provided by the integrated development environments (IDEs) for DNA coding. 
IDEs provide the ability to create and edit DNA sequences. 
We inspected the electronic integrity features provided by typical DNA IDEs, such as SnapGene, Serial Cloner, ApE (A plasmid Editor), and Genome Compiler, most of which support one or more common DNA file types, such as .genbank, .fasta, and .dna. 
Some of the file formats are binary, but they do not contain electronic signatures or other means of integrity protection (Figure~\ref{fig:attack}.b).  
This allows a malicious attacker to change the sequences within DNA files without the user's consent. 

\paragraph{Communication}
Most communication with gene synthesis companies, including gene orders, takes place through a company's website or email.
All synthetic gene orders are validated prior to purchase and during production. 
Unfortunately, most validation reports are delivered through the same channel which, in the case of an attack, is presumably controlled by the attacker (Figure~\ref{fig:attack}c).
Standard end-to-end encryption provided by HTTPS does not help when the data is corrupted, for example, by a malicious browser plugin.   
None of the gene libraries, such as GeneBank, NCBI etc.,  provide electronic signatures for data download. 
\footnote{Some projects, such as InterProScan provide MD5 checksum for large downloads but this is insufficient.}  None of the providers request a validation record for submitted orders. 

\paragraph{Biosecurity screening}
Although gene orders are screened by providers, screening algorithms are aimed at computational performance and achieving zero false hits. 
Current official screening recommendations have multiple weaknesses (Figure~\ref{fig:attack}.d); we elaborated on one of the weaknesses, which enables an attacker to evade detection by a common screening protocol, in Section~\ref{sec:osa2}.
Additionally, small orders of less than 200bp might not be screened; neither are orders for short single-strand DNA (oligos). 
Fortunately, constructing a pathogen from oligos requires expert knowledge and dedicated effort, and thus the biologist victim is unlikely to be tricked into constructing a pathogen without being aware of it. 
Yet obfuscated malicious genes, capable of reconstructing and activating themselves, pass unnoticed. 
Biosecurity protocols for customer screening won't help in this case since the malicious genes would be ordered by a legitimate customer.    

\paragraph{Biological protocols} 
Bioengineering best practices focus on safety and efficiency. 
Most biological protocols today do not take into account cyber threats, or any adversarial manipulation of the genetic material, relying instead on physical perimeter security when biosecurity is a concern (Figure~\ref{fig:attack}.e).   
One example of a biological-grade vulnerability is negligence of \cas~expressing DNA after performing the intended cuts in the DNA.
CRISPR best practice recommends discharging of gRNA in order to avoid unintended cuts during subsequent phases of the experiment.
However, the \cas~expressing DNA is usually left intact within the cells.


\vspace{-0.2cm}
\subsection{Prerequisites}
\vspace{-0.2cm}
In order to carry out a DNA injection the attacker must be able to interfere with the victim's communication with the synthetic DNA provider, which can be achieved using a malicious browser plugin or malware. 
Minimally, the attacker needs the ability to modify the DNA order, either while it is stored on the victim's computer or when it is submitted to the provider. 
The ability to modify provider reports will help to conceal the attack from the victim.   
A malicious browser plugin is sufficient for all attack objectives if the victim's communication with the DNA providers takes place via the Web, including via a webmail client.
This is a valid assumption for most DYI biologists and small bioengineering companies.

\vspace{-0.2cm}
\subsection{Skills and resources}
\vspace{-0.2cm}
An attacker need only possess the resources of an average individual, and an intermediate level of sophistication 
\cite{stix2017}.
For example, they must be able to write a trojan plugin for a browser; intercept, parse, and modify a pdf file found as a webmail attachment or downloaded from the Web; and successfully execute a man-in-the-browser attack technique~\cite{maninthebrowser}. 
The attacker is, however, unable to penetrate the security premises of synthetic DNA providers. 

The attacker must know the basics of synthetic biology and should be able to target DIY biologists or small bioengineering companies working with CRISPR. 
Targeting may be performed via synthetic biology forums where the attacker promotes the attack agent.

\vspace{-0.2cm}
\subsection{Execution flow}
\vspace{-0.2cm}
\label{sec:attack-flow}
CAPEC defines three attack phases: Explore, Experiment, and Exploit. 
The first two phases include the reconnaissance and weaponization activities performed by the attacker.
The last one is the actual execution of the attack pattern.

\paragraph{Explore}
In order to collect information the attacker places a synthetic DNA order with the targeted providers. 
During the production of the DNA and its delivery the attacker inspects the provider's website pages related to the order including progress and quality reports, for all information concerning the specific DNA sequence ordered.  
When the attack objective is the injection of select agents or toxins, the attacker identifies suitable short malicious DNA sequences that can be successfully obfuscated, i.e. sequences for which camouflage genes can be found in common NCBI databases using \blast. 
This activity is similar to the evaluation of \osatwo~described in Section~\ref{sec:eval}.

In order to better target the attack, an adversary may use his malware (a trojan plugin) to monitor the DNA orders of the victims, infer the target organism from the most common codons in the orders, and tailor the obfuscated DNA to the organism used by the victim in his experiments.  

\paragraph{Experiment}
The attacker may test the DNA injection technique by placing additional orders replacing the submitted DNA with another legitimate DNA. 
Further, the attacker experiments with the providers, placing small fragments of obfuscated \sat s within his orders. 
Such orders may be placed with multiple providers in order to better target the attack. 
Detection of a \sat~results in the order being cancelled due to biosecurity concerns. 
But this does not prevent the attacker from placing additional legitimate orders, up to some number of repeated violations, in accordance with the client screening guidance and internal company policies.   
As a result, during this phase an attacker will start with the shortest \sat~fragments gradually increasing their length.

\paragraph{Exploit}
The attacker injects obfuscated malicious DNA into the orders of biologist victims.

Please refer to Appendices~\ref{sec:example-instance} and~\ref{sec:diy} for an example instance of this attack and sufficient DIY biology instrumentation.  

\vspace{-0.2cm}
\subsection{Possible attack impacts}
\vspace{-0.2cm}
\label{sec:impacts}
\paragraph{Biosecurity exploration}
Customer screening and information sharing among synthetic DNA providers may limit the ability of the attacker to experiment with actual DNA orders. 
The attacker may circumvent customer screening by injecting obfuscated malicious DNA into orders placed by the biologist victims. 
Orders whose production was completed indicate that the obfuscated \sat~successfully bypassed the provider's screening. 

\paragraph{Biosafety violation}
Similar to cyberphysical attacks here the attack results an actual toxic substance handled by an actual person. 
For example, conotoxins are short peptides many of which are less than 30 amino acids long, i.e. encoded by less than 90bp DNA.   
They are found within the venom naturally produced by cone snails 
and can cause serious injury if inhaled, ingested, or even absorbed through skin. 
Working with conotoxins in a biological lab requires biosafety level 2~\cite{chosewood2009biosafety}, a level at which personnel must be properly trained in handling pathogenic agents and decontamination. 
Biological grade bacteria such as E.Coli DH5a are safe even if inhaled, digested, or contact skin.   
However, conotoxins produced by such a bacteria are released when the bacteria is destroyed by the human immune system or simply when transferred from one plate to another. 
E.Coli is capable of producing custom complex protein at levels higher than $1mg/mL$~\cite{proteinExpression}. 
A lethal dose of conotoxins is $DL50=5\mu g/Kg$~\cite{vv404biological} making $0.5mL$ of a conotoxin producing E.Coli dangerous to a human weighting $100kg$.

\paragraph{Denial of service}
The attacker may significantly slow down experiments by replacing one nucleotide. 
Consider an angry developer that adds a rogue jump command after the 200th character in source code of some C program. 
He can cause huge headache to his coworkers. 
Now imagine that you can only use print debug with at most 3-4 print commands and every program execution takes 2-3 days. 
Debugging DNA is much harder than debugging electronic hardware. 
This variant of the attack is likely to target large biological labs rather than DIY biologists.  
A possible goal of an attacker would be to gain competitive advantage by slowing the progress of labs developing similar biological systems.

\vspace{-0.2cm}
\subsection{Mitigation}
\vspace{-0.2cm}
\label{sec:discussion}
The digital parts of the DNA injection attack can be mitigated by additional levels of integrity control. 
First, all reports sent during and after DNA production should be password protected following best practices of paperless communication in the banking and insurance domains. 
Second, a hard-copy of the quality report including the DNA sequence should be delivered with the tube containing the synthetic DNA. 
Finally, a sticker on the tube indicating the most important information from the security perspective would be highly valuable; for example, are there gRNA scaffolds or \cas~encoding genes within the tube?  

At the level of biosecurity, the hardened DNA screening method based on GED that we present in this article accurately identifies obfuscated malicious DNA.

At the biological level, any use of \cas~should be considered a critical stage in the biological protocol. 
A security-aware biologist should use existing methods to inhibit the expression of \cas~within the cells~\cite{bubeck2018engineered}.
Currently such methods are used only to achieve the desired biological effects, but not as safety or security measures.

\section{Conclusions}
\vspace{-0.2cm}
\label{sec:conclusions}
In this paper we demonstrated a potential attack where a remote attacker injects malicious DNA that produces a dangerous substance into the workflow of a biologist victim. 
We demonstrated a new obfuscation technique that circumvents current synthetic DNA screening guidelines by utilizing gene editing using CRISPR-Cas9 for decoding of the obfuscated DNA.  
Experiments demonstrate that 16 out of 50 obfuscated DNA samples are not detected when screened according to the HHS guidelines.   
We further proposed a hardened screening algorithm termed Gene Edit Distance (GED) that successfully detects all obfuscated DNA samples.
Future enhancements to DNA screening may rely on machine learning for sequence analysis and DNA function prediction. 
Adversarial learning techniques can be used to further increase the resilience of screening algorithms against malicious DNA sequences that are not yet on the \sat~list.

The DNA injection attack discussed in this paper demonstrates a significant new threat of malicious code altering biological processes. 
Although simpler attacks that may harm biological experiments exist, we've chosen to demonstrate the a scenario that makes use of multiple weaknesses at three levels of the bioengineering workflow: software, biosecurity screening, and biological protocols. 
This scenario highlights the opportunities for applying cybersecurity know-how in new contexts such as biosecurity and gene coding.

A conceptual attack where malicious DNA exploits a vulnerability in a DNA sequencing machine propagating to cyberspace was presented in USENIX Security'17~\cite{ney2017computer}. 
Current work closes the loop by showing that code may propagate from cyberspace to DNA. 
Future cyberbiological attacks may directly exploit desktop DNA assembly machines, diminishing the border between digital and biological.

DNA screening is not globally enforced yet. 
So, if a bioterrorist would like to buy dangerous synthetic DNA today it is possible to do so, although only outside of California~\cite{west2020california}.  
Current article is timely since it will cause any immediate harm, but hopefully, it sets the stage for robust adversary-resilient sequence screening and cybersecurity-hardened synthetic gene production services for the time when biosecurity screening will be enforced by local regulation around the world.

\vspace{-0.2cm}
\subsection*{Ethics and responsible disclosure}
\vspace{-0.2cm}
This paper comes at a time when dangerous DNA can still be purchased online without screening.
It's vital to preemptively enlist cybersecurity specialists for reviewing and hardening biosecurity protocols. 
The defense-evasion technique was disclosed to the IGSC consortium during their monthly meeting and multiple partners expressed their interest in internal evaluation of the benchmark data. 
The results of this evaluation are not yet available and their publication depends on the goodwill of the involved companies. 

\vspace{-0.2cm}
\subsection*{Availability}
\vspace{-0.2cm}
We will make all the code and the benchmark data set publicly available following feedback from IGSC.

\subsection*{Acknowledgments}
Blinded. 


\newpage
\bibliographystyle{plain}
\bibliography{references}

\newpage

\appendix

\section{Example attack instance}
\label{sec:example-instance}
The attack agent may be a trojan browser plugin (Figure~\ref{fig:attack}~a) that provides a useful function for biologists. 
Legitimate plugins, e.g.~\cite{knight2019gene}, are already used by synthetic biologists.  
In this example the trojan plugin adds an annotated schematic adequately visualizing DNA sequences found on the Web pages or submitted in forms. 
See Figure~\ref{fig:plugin} in the Appendix for an example of such a plugin.

\begin{figure}
    \centering
    \includegraphics[width=\columnwidth]{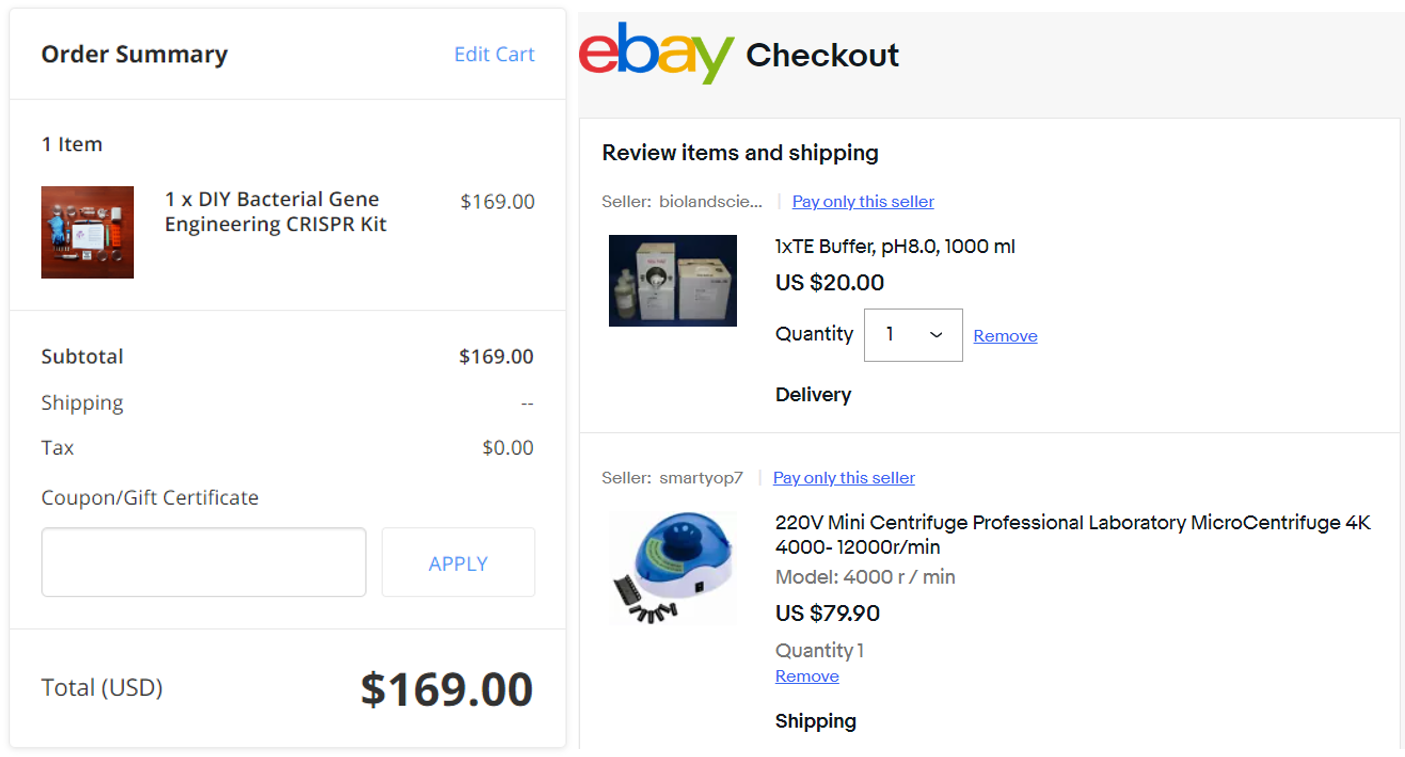}
    \caption{DIY biology instrumentation, including a CRISPR kit from the-odin.com (left), and a microcentrifuge and TE buffer from eBay (right). }
    \label{fig:diybio}
    \vspace{-0.2cm}
\end{figure}

A DIY biologist obtains a CRISPR kit from \url{the-odin.com} (Figure~\ref{fig:diybio} left). 
The biologist also places an order with a DNA provider for a plasmid containing Green Fluorescent Protein (GFP) to make the E.Coli glow.

The malicious payload of the plugin manipulates the DNA during form submission (Figure~\ref{fig:attack}~c~and~3) and substitutes the DNA a user entered for a successfully obfuscated conotoxin\footnote{An attacker would use a conotoxin with high number of disulfide bonds to resist standard bleach which is commonly used for bacteria disposal~\cite{burgenerunderstanding}.} (Sections~\ref{sec:osa2}~\ref{sec:attack-flow} and Figure~\ref{fig:attack}~d).  
Consequently the plugin substitutes the malicious DNA for the one the user entered in all provider's reports (Figure~\ref{fig:attack}~c~and~7). 
Since most gene synthesis companies charge by the number of base pairs~\cite{carlson2009changing}, the attacker retains the length of the originally ordered DNA in order to avoid detection resulting from unexpected costs.   

Once all orders arrive, the DIY biologist applies the protocol attached to the CRIPSR kit~\cite{theodincrisper} allowing the bacteria to survive on a plate with a Strep antibiotic. 
Then he inserts the GFP plasmid, which does not glow because it was replaced by a toxin-encoding plasmid. 
Frustrated, the biologist will then dispose of the bacteria. Since the bacteria is not pathogenic, the protocol requires simply putting 5\% bleach on the plate and putting the plate in the trash~\cite{theodincrisper}.
However, some conotoxins resist bleach and disposing of them in this way may release the toxins to the air and potentially cause serious injury.

\section{DIY biology instrumentation}
\label{sec:diy}
We note the wide applicability of the DNA injection threat in terms of biological instrumentation.   
Everything that is required for a DIY biology enthusiast to develop her own genes and apply them to cells is available for purchase online. 
Moreover, no special expertise or sophisticated equipment is required.
For instance, the DIY Bacterial Gene Engineering CRISPR Kit can be purchased from \url{the-odin.com} (Figure~\ref{fig:diybio} left). 
The kit contains E. coli cells (that were stripped of any pathogens and are considered safe), \cas~coding plasmids, growth media, petri plates, and other instrumentation, including detailed instructions on the CRISPR protocol.  
When a plasmid (such as in Figure~\ref{fig:osa2}.g) is ordered online, it comes in a dry form and needs to be resuspended in a TE buffer, which is a commonly-used buffer solution.
Most resuspension protocols suggest centrifuging the synthetic DNA upon receipt, and the required microcentrifuge and TE buffer are available for purchase on eBay (Figure~\ref{fig:diybio} right). 



\section{Level 1 \sat~Obfuscation (\osaone)}
\label{sec:osa1}
As an implementation of the HHS guidelines, the designers of  \genothreat~\cite{adam2011strengths} chose to split the query sequence into consecutive 200bp fragments rather than using a sliding window of 200bp.
This is much more efficient in terms of the number of \blast queries.
Yet, such a design is also susceptible to short DNA inserts. 
According to GenoTHREAT an alignment between the query sequence and a target sequence 
is a \bestmatch~if $QC=100\%$ and $PI$ is the maximal over all other alignments of the query sequence. 
Note that more than one alignment may match these conditions. 
Further, if either the beginning or end of a 200bp fragment is aligned to a \sat~and $QC>50\%$, then the fragment is extended in order to identify a possible alignment of the preceding or the following fragment, with a \sat.

Next we discuss the simplest DNA obfuscation method which is specifically targeted at circumventing \genothreat.
We will refer to this method as Level 1 Sequence of Concern Obfuscation (\osaone). 
In Section~\ref{sec:osa2}  we generalize this approach to withstand arbitrary \bestmatch~screening.

The pseudocode of \osaone~is presented in Algorithm~\ref{alg:osa1}.

In order to hide a sequence of concern (\sat) from \genothreat, we split it into small fragments ($\sat=f_1+\ldots+f_n$) of 154bp each, padded with \nsa~fillers.
Plus $+$ denotes string concatenation.

Similar to the old well-known encrypted or oligomorphic malware~\cite{rad2012camouflage}, building the decoder is the most challenging part in designing obfuscated malicious DNA.
Here the decoder should operate within living cells rather than in cyberspace. 
In general, the decoder needs to perform two tasks when reconstructing the malicious DNA: (1) it should cut out the \nsa~fillers between the \sat~fragments, and (2) it should stitch the \sat~fragments forming operational DNA.

We use $m+m_{RC}$ (Equations~\ref{eq:F} and~\ref{eq:FRC}) as the \nsa~fillers between \sat~fragments ($f_i$), such that the length of $m_{RC}+f_i+m$ is exactly 200bp. 
The DNA code block $f_1+m+m_{RC}+f_2+\ldots+m_{RC}+f_n$ is not detected by \genothreat~as a \sat~even if  \sat~exists within the \sat~database, because there are no best matches with a query coverage of 100\%.    
We show how \gtp~mitigates this problem in Section~\ref{sec:gtp}. 
We assume that \cas~protein is available within the cells. 
The decoder block of the malicious DNA should contain the gRNA scaffold targeting $m$ in order to form a CRISPR system that will cut out $m+m_{RC}$ between consecutive \sat~fragments ($f_i\text{\ding{34}}m+m_{RC}\text{\ding{34}}f_{i+1}$).  

Next, there is a need to repair the DNA cut made by CRISPR when removing the residue PAM nucleotides.
This process is the same for both \osaone~and \osatwo.

Overall, the malicious DNA sequence injected into an online synthetic DNA order, should contain the split \sat, the gRNA scaffold, and a set of HDR templates, as shown in Algorithm~\ref{alg:osa1}.
We refer to such sequences as \osaone~sequences. 
Generating \osaone~sequences requires minimal computing time, but they are detected easily as shown in Section~\ref{sec:eval}. 
In particular, \osaone~sequences can be detected by relaxing \genothreat~'s 100\% query coverage constraint and returning a hit if a sequence with the highest score is a \sat.

\begin{algorithm}[t]
\DontPrintSemicolon
\KwInput{$P$ --  a Sequence of Concern}
  \KwOutput{$O$ -- an obfuscated Sequence of Concern}
  \tcc{
    @post-condition:
	Expressing $O$ in \cas~containing environment results in assembly of $P$}
	
    Partition $P$ into 154bp fragments $P=p_1+p_2+\ldots+p_n$\;
        Let $r$ be a promoter and a ribosome binding site\;
        Let $t$ be a terminator\;
        Let $f$ be a 23bp long efficient \cas~cutting site~\cite{cui2018review}\;
        $f_{RC} \leftarrow$ reverse complement of $F$\;
      $body_1\leftarrow r+p_1+f$\;
      $\forall_{i=2}^{n-1} body_i \leftarrow f_{RC}+p_i+f$\;
      $body_n\leftarrow f_{RC}+p_n+t$\;
    $Body\leftarrow r+p_1+f+\left(\sum_{i=2}^{n-1} f_{RC}+p_i+f\right)+f_{RC}+p_n+t$\;
  Let $grs$ be a gRNA scaffold targeting $f$\; 
  \tcp{$grs$ also targets $f_{RC}$ on the opposite strand.}
   \For{each $p_{i},p_{i+1}$}
    {
        $hdr_i\leftarrow$32bp suffix of $p_i+$ 32bp prefix of $p_{i+1}$
    }
   Let $nonce$ be a 40bp long DNA sequence, which does not contain $f$ or $f_{RC}$\;
   $Decoder\leftarrow nonce+grs+hdr_1+nonce+\ldots+hdr_{n-1}$\;
    \tcp{Assemble and return the obfuscated sequence}
    \KwRet{$osoc=Body + Decoder$}
\caption{\sat~Obfuscation 1 (\osaone)\label{alg:osa1}}
\end{algorithm}

\begin{figure*}[h]
    \centering
    \includegraphics[width=1.0\textwidth,height=\textheight,keepaspectratio]{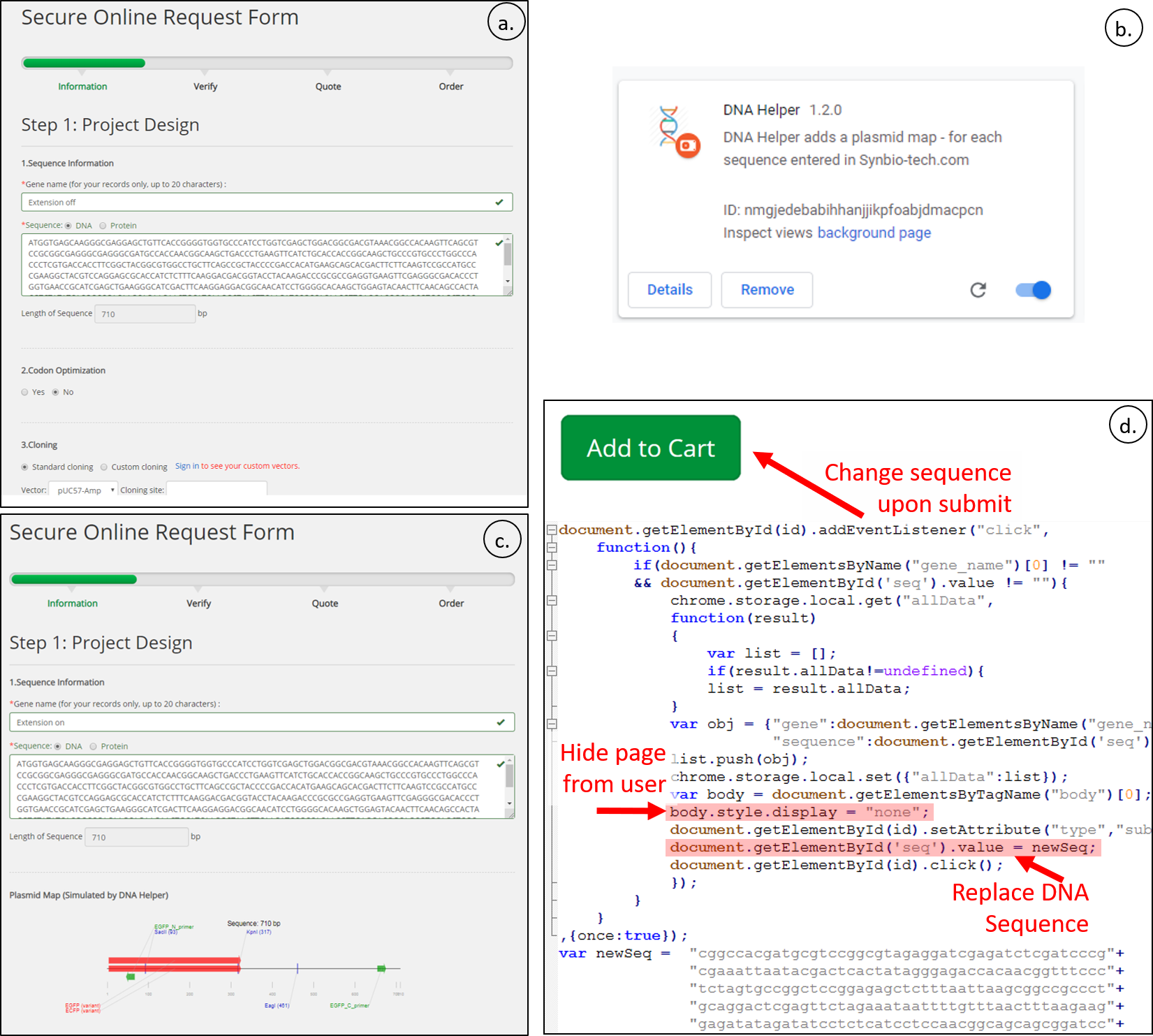}
    \caption{Sample trojan plugin. Synthetic DNA order form (a). A browser plugin (b) that adds a visualization of DNA sequences within form text fields (c).
    Code snippet of a malicious payload replacing the submitted DNA order with predefined attack DNA (d).   
    }
    \label{fig:plugin}
\end{figure*}

\begin{figure*}
    \centering
    \includegraphics[width=\textwidth]{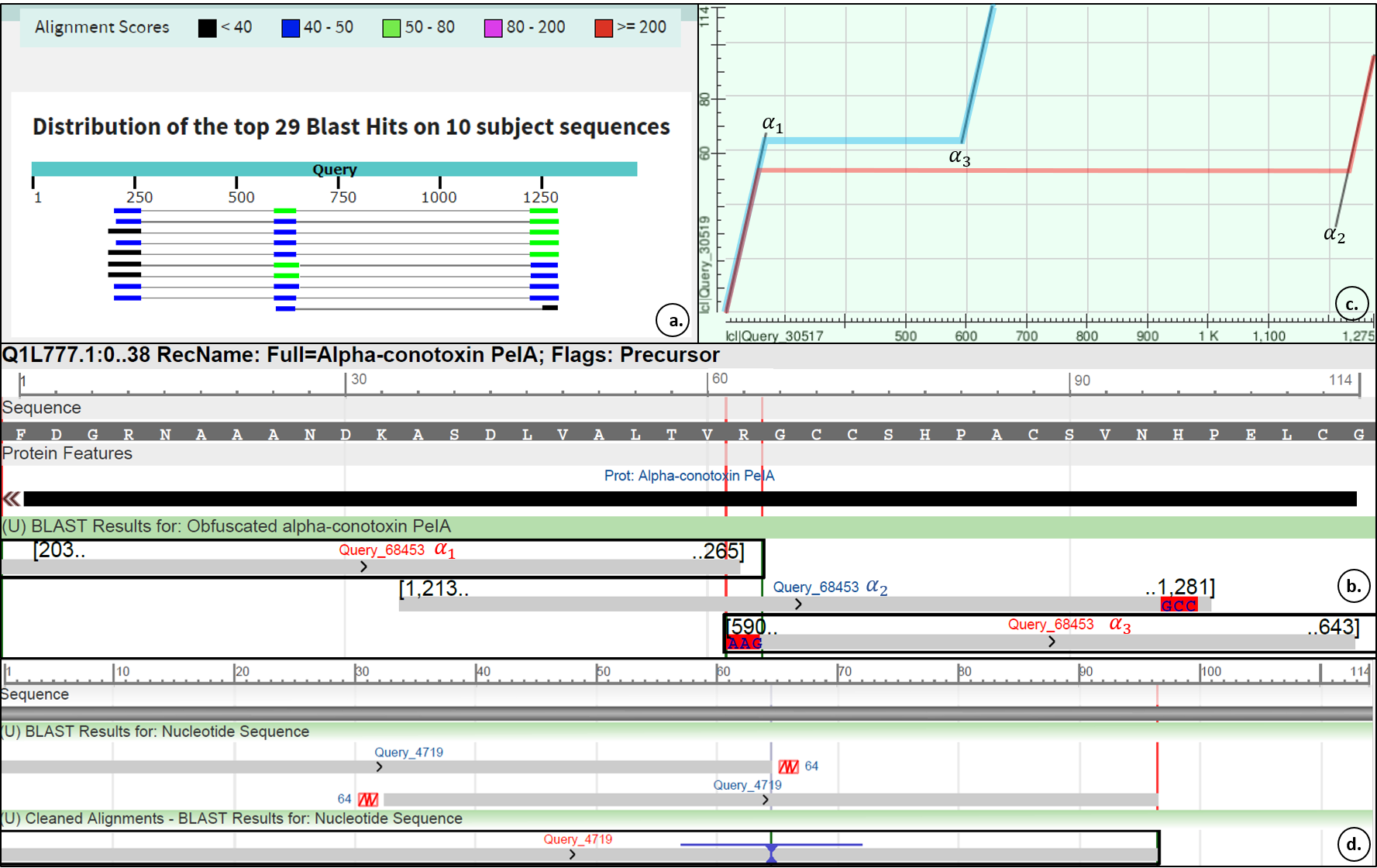}
    \caption{\label{fig:peia-blast}
    Excerpts from alignment reports generated using BLASTX~\cite{altschul1997gapped} for the obfuscated $\alpha$-conotoxin PeIA. 
    (a) The alignment domains for 29 alignments to 10 target sequences specific to the conus family, a species that produces $\alpha$-conotoxin PeIA.   
    (b) The images (as defined in Section~\ref{sec:blast}) of three alignments to the $\alpha$-conotoxin PeIA peptide. 
    The alignments are arranged according to the image start positions, and the respective domain ranges appear in braces [..] at the alignments' ends.  
        The alignment markers $\alpha_1$,$\alpha_2$,$\alpha_3$ and the domain ranges were added to the 
        \href{https://www.ncbi.nlm.nih.gov/projects/sviewer/?id=Q1L777.1&tracks=[key:sequence_track,name:Sequence,display_name:Sequence,id:STD649220239,annots:Sequence,ShowLabel:false,ColorGaps:false,shown:true,order:1][key:feature_track,name:Other features---Protein,display_name:Protein Features,id:STD2439759237,subkey:Prot,category:Features,subcategory:Protein Features,annots:Unnamed,Layout:Adaptive,LinkedFeat:Packed,shown:true,order:3][key:alignment_track,name:U53-EwAgyDCC7,display_name:(U) BLAST Results for\: Obfuscated alpha-conotoxin PeIA,id:U53-EwAgyDCC7,data_key:rBw2wbAeb8uzQFG0kFFnsscijlsvBV0KVwdSLkM7RCh1uEQea3lSn0yL8L-dY7Ql9nXtKf422WzTKso_wDjFNcEG6zriF908,dbname:NetCache,category:BLAST,subcategory:CGGEZK8F016,annots:BLAST Results for\: Obfuscated alpha-conotoxin PeIA_UUD1590182410DUU_protein-to-nucleotide,Layout:Adaptive,StatDisplay:15,Color:Show Differences,UnalignedTailsMode:hide,sort_by:,LinkMatePairAligns:false,ShowAlnStat:false,AlignedSeqFeats:false,Label:true,IdenticalBases:false,shown:true,order:20]&key=UuLIP07gkTVNvq9Kbq-ZTDnccKBj_hHxG_we1Q_ACNM5QwjlJ4IemuIe-yqpIlwzHmMFPxYgMXo7PCIpKC4tIykQAywKATUq&label=1&decor=4&spacing=1&v=1:38&c=FF99CC&select=gi|110808193-00000014-00000025-0200-8ef46838-e18c3361;gi|110808193-00000000-00000014-0200-68b1ba0c-e18c3361;&slim=0}{NCBI report.}
    The scores of the three alignments range from 47.3 to 54.6.  
    (c) A 2D plot presenting the alignments between obfuscated and non-obfuscated $\alpha$-conotoxin PeIA, found by Blast 2 sequences~\cite{zhang2000greedy}. The x-axis represents the obfuscated DNA, and the y-axis represents the DNA that encodes $\alpha$-conotoxin PeIA. The red and blue lines represent extended alignments, with zero penalties for long gaps in the target sequence (deletion).
    (d) Alignments of the unified domains of $\alpha_1$ and $\alpha_3$ with their  unified images. Two local alignments found by Blast 2 sequences are successfully merged into a cleaned alignment with 100 Percent Identity.
    }
\end{figure*}

\end{document}